\documentclass[sigconf,nonacm,screen]{acmart}

\usepackage[shortlabels]{enumitem}
\usepackage{balance}
\usepackage{draftwatermark}

\SetWatermarkText{Preprint}
\SetWatermarkColor[gray]{0.9}
\SetWatermarkFontSize{3cm}
\SetWatermarkAngle{30}
\SetWatermarkHorCenter{11cm}

\newcommand\blfootnote[1]{%
  \begingroup
  \renewcommand\thefootnote{}\footnote{#1}%
  \addtocounter{footnote}{-1}%
  \endgroup
}

\DeclareRobustCommand\ECJlongname{}
\DeclareRobustCommand\EUlongname{}
\DeclareRobustCommand\EClongname{}
\DeclareRobustCommand\DGMOVElongname{}
\DeclareRobustCommand\IVASSlongname{}
\DeclareRobustCommand\UNARlongname{}
\DeclareRobustCommand\ASGIlongname{}
\DeclareRobustCommand\UNRAElongname{}

\copyrightyear{2021}
\acmYear{2021}
\setcopyright{acmlicensed}
\acmConference[AIES '21] {Proceedings of the 2021 AAAI/ACM Conference on AI, Ethics, and Society}{May 19--21, 2021}{Virtual Event, USA.}
\acmBooktitle{Proceedings of the 2021 AAAI/ACM Conference on AI, Ethics, and Society (AIES '21), May 19--21, 2021, Virtual Event, USA}
\acmPrice{15.00}
\acmDOI{10.1145/3461702.3462569}
\acmISBN{978-1-4503-8473-5/21/05}

\begin{document}
\fancyhead{}

\title[Algorithmic Audit of Italian Car Insurance]{Algorithmic Audit of Italian Car Insurance: \\  Evidence of Unfairness in Access and Pricing}

\author{Alessandro Fabris}
\email{fabrisal@dei.unipd.it}
\affiliation{%
  \institution{University of Padua}
  \country{Italy}
}

\author{Alan Mishler}
\email{amishler@andrew.cmu.edu}
\affiliation{%
  \institution{Carnegie Mellon University}
  \country{USA}
}

\author{Stefano Gottardi}
\email{stefano.gottardi@studenti.unipd.it}
\affiliation{%
  \institution{University of Padua}
  \country{Italy}
}
\author{Mattia Carletti}
\email{mattia.carletti@unipd.it}
\affiliation{%
  \institution{University of Padua}
  \country{Italy}
}

\author{Matteo Daicampi}
\email{daicampi.matteo@spes.uniud.it}
\affiliation{%
  \institution{University of Udine}
  \country{Italy}
}

\author{Gian Antonio Susto}
\email{sustogia@dei.unipd.it}
\affiliation{%
  \institution{University of Padua}
  \country{Italy}
}

\author{Gianmaria Silvello}
\email{silvello@dei.unipd.it}
\affiliation{%
  \institution{University of Padua}
  \country{Italy}
}

\renewcommand{\shortauthors}{Fabris et al.}

\begin{abstract}
  We conduct an audit of pricing algorithms employed by companies in the Italian car insurance industry, primarily by gathering quotes through a popular comparison website. While acknowledging the complexity of the industry, we find evidence of several problematic practices. We show that birthplace and gender have a direct and sizeable impact on the prices quoted to drivers, despite national and international regulations against their use. Birthplace, in particular, is used quite frequently to the disadvantage of foreign-born drivers and drivers born in certain Italian cities. In extreme cases, a driver born in Laos may be charged 1,000€ more than a driver born in Milan, all else being equal. For a subset of our sample, we collect quotes directly on a company website, where the direct influence of gender and birthplace is confirmed. Finally, we find that drivers with riskier profiles tend to see fewer quotes in the aggregator result pages, substantiating concerns of differential treatment raised in the past by Italian insurance regulators.  
\end{abstract}





\maketitle

\section{Introduction}
\blfootnote{Preprint of the paper published in the 4th AAAI/ACM conference on AIES (2021)}

Car ownership is an important factor for employment and, more broadly, for participation in the economic, social and political organization of many societies \cite{ong2002:co,smart2020:dr}. This may be especially true in Italy, second only to Luxembourg for car ownership among EU states \cite{eudg_transport2019}. Auto insurance makes vehicle ownership and usage less hazardous from an individual financial perspective  \citep{frezal2019:fu}. It acts as a risk-pooling device, covering drivers against liability for bodily injury and property damage in exchange for a premium. Companies are developing increasingly complex machine learning-based models for car insurance pricing \citep{chapados2002:ec, yan2020:ru}.

The legal liability connected to driving a vehicle in Italy, and the car insurance system itself, are known as the Motor Vehicle Liability system (\emph{Responsabilità Civile Autoveicoli} - RCA). It is mandatory to purchase RCA coverage before using or keeping a motor vehicle on Italian public roads. RCA is regulated by the national Institute for the Supervision of Insurance (\emph{Istituto Per la Vigilanza Sulle Assicurazioni} - IVASS), which oversees the industry, protecting customers and ensuring transparency while also promoting market stability and the financial viability of businesses. 

Over the last decade, the use of sensitive features such as gender and nationality in the Italian car insurance industry has been regulated. The European Union has adopted legislation which prohibits the direct use of gender for setting insurance premiums \cite{ecj2011, eu2012:guidelines}. After finding evidence of discrimination in pricing on the basis of nationality, IVASS and the National Anti-Racial Discrimination Office (\emph{Ufficio Nazionale Antidiscriminazioni Razziali} - UNAR)  issued soft regulation that encourages companies to avoid using nationality-related factors, such as birthplace and citizenship, as inputs to risk models \cite{unar2012, ivass2014}. 

Concurrently, comparison websites (also called aggregators) have become a primary point of access to RCA subscription, claiming half of the total gross written premiums in the Italian vehicle insurance market in 2017 \cite{mckinsey2018}. Due to their growing importance, comparison websites have come under increased scrutiny and regulation. In a past investigation on RCA aggregators \cite{ivass2014:aggregator}, IVASS found anecdotal evidence of \emph{output variability} in connection with risk profile: result pages from comparison websites seemed to display fewer quotes to driver profiles associated with higher risk. While the evidence, based on a limited sample of 7 driver profiles, was not conclusive, this was highlighted as a potential problem of differential treatment, providing uneven opportunities to different driver segments. 

To the best of our knowledge, no study to date has analyzed the direct influence of gender on car insurance pricing in Italy, despite the laws and regulations that limit its use. Additionally, we are unaware of any study that has examined the influence of nationality-related features on pricing since regulations were issued by IVASS and UNAR. Finally, we are unaware of a systematic study of output variability in RCA comparison websites. 

\noindent \textbf{Research questions.} In this paper we exploit nearly 20,000 quotes gathered on a popular comparison website to audit output variability in the aggregator's result pages, along with the RCA pricing practices of nine companies, representing a significant share of the market. More specifically, we ask:

\begin{enumerate} [start=1,label={\bfseries RQ\arabic*:}, leftmargin=1cm]
\item What are the factors that play a major role in setting RCA premiums?
\item Do gender and birthplace directly influence quoted premiums?
\item Do riskier driver profiles see fewer quotes on comparison websites?
\end{enumerate}

We gather our data under a full factorial design, in which we vary some characteristics of the ``applicant'', while holding the remaining characteristics fixed. 
Additionally, we collect data directly from a single RCA company's website in order 
to verify whether the key trends detected on the aggregator occur at the company level.

\noindent \textbf{Contributions and outline.} We discuss the relevant background and normative fairness framework in Section \ref{sec:background}. Section \ref{sec:data} describes the Design of the experiment and the resulting dataset. In Section \ref{sec:overview} we address \textbf{RQ1}, finding that age, residence, claim history and insured vehicle type all have a substantial influence on premiums. In section \ref{sec:discr}, we address \textbf{RQ2}, showing that both gender and birthplace have a direct impact on quoted RCA prices. Birthplace is especially problematic from a fairness perspective, as we find a consistent financial disadvantage for foreign-born drivers and for drivers born in certain Italian cities. In Section \ref{sec:target}, we address \textbf{RQ3}. We find that the variability in the number of quotes displayed by the aggregator seems aligned with a deliberate choice: profiles perceived as risky tend to see fewer quotes, raising concerns of transparency and unequal opportunity. 
In Section \ref{sec:aggregator_influence} we verify that the pricing trends measured on the aggregator are confirmed on a company's website. Section \ref{sec:concl} summarizes our conclusions while outlining the limitations of this paper and directions for future work. 

\section{Background and Related Work}
\label{sec:background}

\subsection{Protected attributes and fairness criteria}
\label{sec:normative}

We focus on gender and birthplace as sensitive features both (1) because there exists legislation regulating their use for insurance pricing in Italy, and (2) because RCA websites require users to input these features before generating quotes. Other features that are commonly invoked in studies of fairness and discrimination are either not collected by insurance websites (e.g. race and ethnicity) or are currently permitted under the law as inputs to risk models (e.g. age).

Gender is often conflated with sex in European legislation on insurance \cite{eu2012:charter, ecj2011, eu2012:guidelines}. The forms in the websites we crawled prompt ``The driver is'', providing the options ``female'' and ``male''. We refer to this feature as \emph{gender} throughout the manuscript and follow the binary framing common in the industry and current legislation. 

The principle of gender equality is enshrined by Articles 21 and 23 of the Charter of Fundamental Rights of the European Union \cite{eu2000:charter,eu2012:charter}. Gender equality has been explicitly operationalized in the context of insurance, with Article 5(1) of Council Directive 2004/113/EC \cite{eu2004:implementing}, stating that no difference in individuals' premiums can result from the use of gender as an explicit factor, and fully confirmed in a 2011 judgement by the European Court of Justice \cite{ecj2011}.

Official guidelines on the application of the ruling \cite{eu2012:guidelines} explicitly mention motor insurance, clarifying that indirect discrimination remains possible where justifiable: 
``For example, price differentiation based on the size of a car engine in the field of motor insurance should remain possible, even if statistically men drive cars with more powerful engines''.
Moreover, information about gender may still be collected, stored and used, e.g. to monitor portfolio mix or for the purposes of reinsurance. 

Nationality-related features were freely used as inputs to actuarial models in the Italian industry until 2010, when a Tunisian citizen residing in Italy since 1992 sued his car insurance company after being quoted a 30\% surcharge due to his citizenship. The lawsuit was later extended to other companies found to engage in similar practices. Following extensive press coverage and further evidence brought forth by non-Italian citizens, the matter came to the attention of UNAR, who opened an investigation in concert with IVASS and the National Association of Insurance Companies (\emph{Associazione Nazionale fra le Imprese Assicuratrici}). IVASS reported that 25\% of companies in their sample took nationality into account as a risk factor. UNAR contacted companies found to charge foreign-born drivers more than Italian-born drivers; one company clarified that birthplace is intended as a proxy for the country where drivers obtain their license, and that learning to drive under different traffic rules and road signs represents an important risk factor. Based on these circumstances, in light of extensive analysis of national and European anti-discrimination law, UNAR issued a general recommendation to the industry, requesting that companies charge the same premiums to Italian and non-Italian citizens, all else being equal \cite{unar2012}.

Shortly thereafter, two companies under lawsuit issued a press release, stressing the absence of discriminatory intentions in their practices and committing to removing citizenship from the parameters explicitly used in their risk models \cite{asgi2012:press}. The lawsuit was thereby settled and dropped. Finally, in 2014, IVASS issued soft regulation, recalling and echoing the recommendation from UNAR, with wording explicitly focused on \emph{birthplace} \cite{ivass2014}. IVASS invited all insurance companies operating in Italy to ``reconsider this criterion and put in place any activity deemed necessary in order for car insurance quotes and contracts not to take country of birth into account''.
This regulation clarifies unambiguously that birthplace -- not only citizenship -- is a sensitive factor, and that its direct utilization in actuarial models is considered discriminatory by IVASS. Henceforth, we refer to a single nationality-related variable, i.e. \emph{birthplace}, given that this is the information currently queried by websites, and distinguish it from citizenship where relevant for the discussion.

In sum, the regulatory framework against discrimination in car insurance described above, comprising EU legislation on gender and Italian soft regulation on birthplace, permits the collection of protected attributes while forbidding their direct utilization, thereby aligning with the criterion of \emph{Fairness Through Unawareness} \cite{grgic2016:case,kusner2017:cf} defined below.

\noindent \textbf{Definition} (Fairness Through Unawareness - FTU). Consider a function (equivalently, ``algorithm'') $f: \mathcal{S} \times \mathcal{X} \mapsto \mathcal{Y}$, where $\mathcal{S}$ represents a sensitive feature, $\mathcal{X}$ represents additional covariates, and $\mathcal{Y}$ represents an output space. The algorithm satisfies FTU with respect to $S$ so long as $f(s, x) = f(s', x)$ for all $s, s' \in \mathcal{S}$ and $x \in \mathcal{X}$. In other words, the algorithm essentially does not utilize the sensitive feature. If the sensitive feature does not form part of the input to the algorithm, then the algorithm  trivially satisfies FTU with respect to that feature.

Given its alignment with current regulation, we adopt FTU as the relevant criterion for the purposes of our algorithmic audit, while recognizing that other notions of fairness may be salient in different contexts.

\subsection{Comparison websites}
\label{sec:comparison}

Comparison websites act as digital intermediaries between customers and insurance providers, typically charging the latter a commission while providing a free service to the former \cite{ivass2014:aggregator,mckinsey2018}. Their penetration in the European car insurance market has increased dramatically over the last decade. Focusing on the Italian market, in 2017 aggregators reached a 48\% share of the total motor gross written premiums \cite{mckinsey2018}. Beyond their importance for direct sales as insurance brokers, comparison websites provide a useful information service for drivers, who can efficiently compare different car insurance options from a single point of access and benefit from increased market transparency.

In a 2014 investigation on comparison websites, IVASS highlighted a few critical aspects \cite{ivass2014:aggregator}. They tested 7 different driver profiles on 6 competing comparison websites, finding anecdotal evidence of \emph{output variability}: result pages seemed to display fewer quotes to driver profiles associated with higher risk. This was stressed by IVASS as a potential problem of differential treatment, providing uneven opportunities to different driver segments. Responding to concerns outlined in the report, some comparison websites provided a technical explanation connected to timeouts.
The IVASS report concluded that it was impossible to ascertain whether indeed the variability in the number of quotes was due to technical reasons or strategic choices.

It is worth noting that Italian law imposes a dual \emph{duty to contract}, which applies to both drivers and insurers. According to Article 132 of the Private Insurance Code \cite{insurancecode:2007}, insurers are obliged to offer RCA coverage to all drivers, regardless of their risk profile. 
Article 132-bis, introduced in 2017, recognizes the growing importance of intermediaries in car insurance, such as brokers and agents, who are required to inform customers exhaustively and transparently with respect to premiums offered by all companies with which they have a broker agreement. 

Our reasons for resorting to a comparison website to acquire car insurance quotes are thus twofold. On one hand, we aim to analyze the direct impact of gender and birthplace on quoted prices in a driver-centric fashion, utilizing this prevalent modality of market access. On the other hand, we are interested in auditing patterns of unequal treatment for different users anecdotally highlighted by IVASS.

\subsection{Algorithmic audit}

An algorithmic audit can be characterized as a study of algorithms, products and services, aimed at uncovering meaningful relationships between inputs and outputs. As automation becomes increasingly embedded in society, processes designed to reverse engineer and uncover key aspects of algorithms and automated decision systems are fundamental. Auditing is a central part of fairness, accountability and transparency, allowing communities to keep technology and decision systems in check and ensure that they are aligned with specific values and requirements.

Among other notable works in this area, researchers have audited personalization in search engines \citep{hannak2013:mp}, price discrimination on e-commerce platforms \citep{hannak2014:mp}, racial bias in judicial risk assessment \citep{skeem2016:rr}, sources of bias in political queries on Twitter \citep{kulshrestha2017:qs}, gender- and race-based differences in accuracy of face analysis technology \citep{buolamwini2018:gs} and radicalization on YouTube \citep{ribeiro2020:ar}.

Several other researchers have investigated discrimination in auto insurance pricing. \citet{harrington1998:rr} used data from Missouri to examine whether insurance profits were higher in ZIP codes with a higher percentage of minorities. They found no evidence of redlining or other racial discrimination. In subsequent work, \citet{ong2007:rr} arrived at a different conclusion. They gathered 836 quotes, varying only the ZIP code while holding all other inputs constant. 
They found that, after accounting for risk factors, socioeconomic factors in a neighborhood such as percentage of poor residents and black residents correlated with higher premiums. This work is the closest to our study, as it is based on quotes gathered with full control of the inputs, some of which are fixed while others are varied according to an experimental design. Most recently, ProPublica analyzed payouts and premiums for car insurance in California, Illinois, Texas and Missouri, coming to similar conclusions that redlining practices affect minority neighbourhoods unfavourably \citep{angwin2017:mn, larson2017:er}.

To the best of our knowledge, no such audit has been conducted for the Italian market; our aim is to close a transparency gap between industry practice and current regulation on the equity of RCA pricing and access.

\section{Data}
\label{sec:data}

Our experimental design and data collection procedure are motivated by the three research questions described in the Introduction. We choose a common strategy to address all three questions: gathering quotes from several companies on a popular RCA comparison website as we vary the drivers' profiles across features that are known \emph{a-priori} to generate sizeable variations in RCA premiums, as detailed in technical reports, trade magazines and domain-specific websites \cite{consumarori2019, ivass2020q2}.

We tried to balance this principle of sizeable output variability with one of sample representativeness. For example, when deciding which vehicles to consider, we restricted our options to the best-selling cars in the Italian market, thus neglecting pricey luxury vehicles which are likely associated with the most expensive RCA quotes, but are also far from a representative choice for the average Italian driver.

\subsection{Design of experiment (DOE)}
\label{sec:doe}
\begin{table}[tb]
  \caption{DOE for data collection.}
  \label{tab:doe}
  \begin{tabular}{lp{3.1cm}l}
    \toprule
    Feature & Values tested & Brief description\\
    \midrule
    gender & F, M & driver's gender\\
    birthplace & Milan, Rome, Naples, \newline Romania, Ghana, Laos & driver's place of birth\\
    age & 18, 25, 32 & driver's age\\
    city & Milan, Rome, Naples & driver's residence\\
    car & OLED, NSEP & insured vehicle type\\
    km\_driven & 10,000, 30,000 & kms driven yearly\\
    class & 0, 4, 9, 14, 18, None (-1) & claim history summary \\
  \bottomrule
\end{tabular}
\end{table}

We define a full factorial experiment, based on protected features (gender and birthplace) 
and features which are widely recognized as significant for pricing
such as driver age, municipality of residence, car, yearly mileage and claim history \citep{cosconati2018:nn, consumarori2019, ivass2020q1}. Table \ref{tab:doe} summarizes our DOE. 

We let \textbf{gender} take the two values permitted in the comparison website: male (M) and female (F). For \textbf{birthplace}, we consider 
Romania, an EU member state with over 1.1 million citizens residing in Italy, along with Ghana and Laos, two countries in completely different geographical areas which also differ greatly for the number of citizens residing in Italy, estimated at 49,543 and 69, respectively \cite{istat2020}.\footnote{While the quoted source reports the number of people with foreign citizenship residing in Italy, the forms in websites we utilized query for their birthplace. Given that Italy has a naturalization rate close to 2\% \cite{eurostat2020}, it seems unlikely that the number of Laos citizens and Laos-born people residing in Italy will differ by orders of magnitude.} 
It is worth noting that most companies are unlikely to have more than a few tens of Laos-born drivers available as data points to infer the ``effect'' of this factor level. For this reason, pricing policies connected with this factor level plausibly stem from subjective (potentially inadvertent) choices rather than statistically significant inference. Along with these countries, we also consider the 3 largest Italian cities in northern (Milan), central (Rome) and southern Italy (Naples).

According to data on recently underwritten RCA contracts \cite{ivass2020q1, ivass2020q2}, most of the \textbf{age}-related premium variability is concentrated in the youngest age groups. The mean price for the youngest bracket (18-24) is nearly double the national average; premiums decrease with age up to the bracket (35-44), where they align with the national average. For this reason, we focus on a young segment of the population, aged 18, 25 and 32, who, as is typical of Italians at their age, have been driving for 0, 7 and 14 years respectively. 

For \textbf{city} of residence, we consider (again)  Milan, Rome and Naples. These are the three largest cities in Italy, and they represent cultural and economic hubs in northern, central, and southern Italy, respectively. Among the ten most populous Italian cities, residents of Naples and Milan pay, on average, the highest and the lowest RCA premiums, respectively \cite{ivass2020q1, ivass2020q2}.

The type of insured \textbf{car} is reported to impact quoted price significantly, with age, engine displacement and fuel system cited by trade magazines as the key factors. 
Among best selling vehicles from 2008 to 2020 \cite{unrae2017,unrae2020}, the most favourable combination for insurance price is achieved by a 2020 Fiat Panda with a 1.2 litre petrol engine
(\underline{n}ew, \underline{s}mall \underline{e}ngine, \underline{p}etrol - abbreviated as NSEP), while the least favourable is a 2008 Fiat Bravo fitted with a 2.0 litre diesel engine
(\underline{o}ld, \underline{l}arge \underline{e}ngine, \underline{d}iesel - OLED). 

Yearly mileage, or \textbf{kilometers driven}, is often cited as an important factor, due to increased time on the road and consequent chance of causing an accident. 
We let this feature take values 10,000 (a common default setting in aggregators) and 30,000.

Finally, Bonus-Malus System (BMS) \textbf{class} \citep{taylor1997:sb} summarizes driver claim history, which is updated yearly. Classes 1 and 18 are the best and worst respectively. New drivers start at class 14, but can alternatively choose to acquire the BMS class of a relative from the same household when purchasing their first car insurance \cite{bersani2007}. Every year, their class improves by 1 if they had no at-fault accidents and increases by 2 otherwise. 
We investigate the full range available for this feature, from class 1 to class 18, including class 4, 9 and 14 as intermediate values. The aggregator distinguishes between class 1 and ``class 1 for one year or more''; we select the latter value and label it ``class 0''. Finally, we also test a profile with no driving record (class None), which should be equivalent to class 14. 

The full factorial design results in 2,592 unique factor level combinations (\emph{profiles}), of which 1/6 are excluded due to inadmissibility: 18-year-olds are not allowed to drive powerful cars (OLED), reducing the size of the experiment to 2,160 profiles. In setting the (constant) values of the remaining features that are not factors in our study, we aimed for plausible values that are compatible with our chosen factor levels. Our subject is employed, single, and has no children. They are the only driver of the insured vehicle, which is used for both work and leisure.

\subsection{Data collection}
\label{sec:data_coll}
\begin{figure}[tb]
    \centering
    \includegraphics[width=0.8\linewidth]{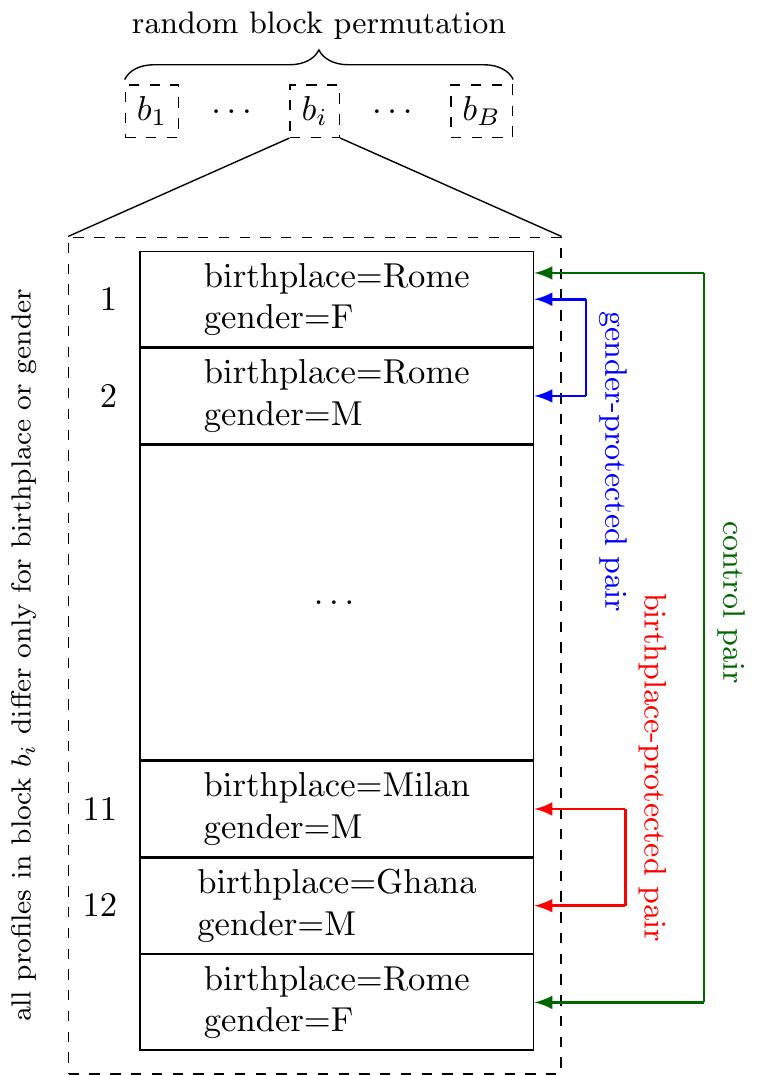}
  \caption{Schematic for our data collection procedure. Quotes for profiles that differ only for birthplace or gender are grouped into $B$ \emph{blocks} of 12 ($B=180$) and collected sequentially in random order. A control quote, identical to the first one, is collected at the end of each block. The blocks are randomly permuted. }
   \label{fig:schematic}
\end{figure}

We gather our data from a famous comparison website, consistently present in the top two search engine results for the query ``comparatore RCA'' (RCA comparison website) and meaningful variations thereof. Insurance groups represented in the search results cover over 60\% of the RCA market. To avoid disrupting the service of the website, we collect fewer than 200 quotes per day during July 2020, over a period of 17 days. We envision 3 plausible sources of disturbance in the pricing signal: (1) the evolution of actuarial models and pricing schemes over time, (2) session duration, with time spent on the website potentially factored into the pricing scheme, and (3) A/B testing on behalf of insurance companies, the comparison website or both. To compensate for these effects we design a doubly-nested randomization with a control group, summarized by Figure \ref{fig:schematic} and described hereafter.

Protected features, likely to cause small fluctuations which we aim to measure carefully, are bundled and rotated. While keeping every other factor constant, we sequentially execute 12 queries, one for each combination of gender and birthplace, normally over a single session, occasionally over two. We call this sequence of 12 queries, identical for every factor but gender and birthplace, a \emph{block}. This is the inner loop, which is randomized so that each combination of gender and birthplace has an equal chance of occurring at any of the 12 slots in the block. Two profiles that differ only for birthplace (gender) make up a birthplace- (gender-) \emph{protected pair}, as exemplified in Figure \ref{fig:schematic}. The remaining features are combined via cartesian product and permuted, thus randomizing the order of blocks. This is the outer loop, comprising $B=180$ blocks in total.

The above procedure should disentangle the features of interest, in particular the protected features, from slow price fluctuations deriving from the evolution of actuarial models and pricing schemes. We further control for unaccounted factors, such as A/B testing or session, by repeating every 12th query. We call these \emph{control} queries, each of them gathered at the end of a block, identical to a regular query gathered at the beginning of the block, with which they form a \emph{control pair}. Overall, the data collection procedure requires the execution of 2,340 queries -- one for each of the 2,160 unique profiles as well as 180 control queries.\footnote{Due to a design flaw, we only executed control queries for the final 71 blocks.} Each query returns between 5 and 12 price quotes, depending on which companies appear in the search results. 

\begin{table}[tb]
\centering
  \caption{Summary of collected insurance quotes.}
  \label{tab:summ_data}
  \begin{tabular}{lrrc}
    \toprule
    Company & Num. Quotes & Frequency & Track\\
    \midrule
    c1/a & 1728 & 80\% & \\
    c1/b & 1728 & 80\% & YES \\
    c1/c & 1152 & 53\% & \\
    c2 & 1787 & 83\% & \\
    c3/a & 1477 & 68\% & \\
    c3/b & 388 & 18\% & YES \\
    c3/c & 690 & 32\% & YES \\
    c4 & 1628 & 75\% & \\
    c5 & 2148 & 99\% & \\
    c6/a & 717 & 33\% & \\
    c6/b & 1428 & 66\% & \\
    c6/c & 360 & 17\% & YES \\
    c7 & 102 & 4\% & \\
    c8 & 2115 & 98\% & \\
    c9 & 2160  & 100\% & \\
  \bottomrule
\end{tabular}
\end{table}

In total, we gather 19,608 yearly quotes from 9 companies (not including control queries), which are summarized in Table \ref{tab:summ_data}. Companies are labeled \texttt{c1} to \texttt{c9}, with arbitrary numbering. Depending on product portfolio and agreements with the comparison website, each company offers up to three different insurance products (labeled `/a', `/b' and `/c'). Products from the same company differ in whether they require a tracking device and whether they include premium services, such as road assistance and coverage of legal expenses. Only one company (c9) provides a quote for every tested profile; two more companies (c8 and c4) appear very frequently (in 98\% and 99\% of the query results, respectively). The remaining companies appear 4-83\% of the time. This is a first hint of output variability, analyzed in Section \ref{sec:target}.

\section{Most Important Factors}
\label{sec:overview}

\subsection{Methods} 
\label{sec:overview_meth}

In this section, we address \textbf{RQ1}, analyzing the average impact each factor has on yearly quoted prices. The comparison website orders quotes for a given profile from cheapest (at the top) to most expensive (at the bottom); hence, we refer to an analysis focused on $k$ cheapest quotes as \texttt{top-k}. We perform the following analyses:

\begin{itemize}
    \item \texttt{top1}: examines the cheapest quote obtained for each profile. This analysis adopts the perspective of a driver solely driven by expense minimization. 
    \item \texttt{top5}: average of the five cheapest quotes obtained for each profile. Average prices correspond to a dual point of view: (1) a driver who is not necessarily seeking the cheapest product; (2) a driver who is ``shopping around'' on the website, comparing several insurance options against their current contract. At least five quotes were returned for each profile.
    \item \texttt{all}: average of all quotes obtained for each profile.
    \item \texttt{c9}: quotes provided by the only company which appeared in result pages for each tested profile, i.e.\  \texttt{c9}. 
\end{itemize}

For each of the four analyses above, we first reduce all the quotes on the result page to a single price, either by selecting the relevant quote (in the \texttt{top1} and \texttt{c9} analyses) or by averaging the selected quotes (in the \texttt{top5} and \texttt{all} analyses). Thus, each profile corresponds to a single price within each analysis.
\begin{figure*}[tb]
    \centering
    \includegraphics[width=\linewidth]{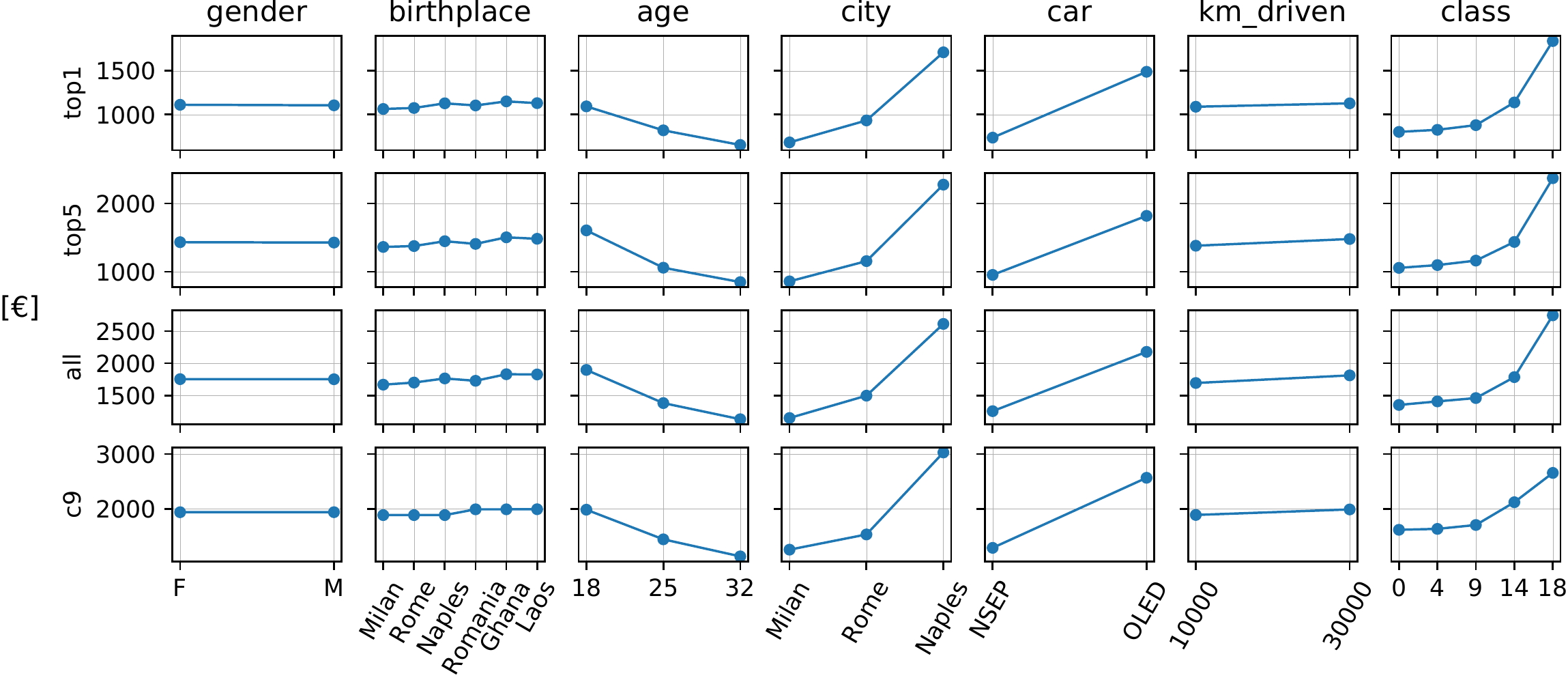}
  \caption{Overview of factor influence on insurance price. Each column represents a different factor, with all its tested levels on the $x$ axis. The $y$ axis depicts the influence of each factor, as a mean price for all profiles with a given factor level, in different analyses: cheapest quote (\texttt{top1} - row 1), average of 5 cheapest quotes (\texttt{top5} - row 2), average of all quotes (\texttt{all} - row 3) and quote by company \texttt{c9} (row 4), the only company present in every result page. Age, city, car and class are confirmed to strongly influence average price, while mileage and protected factors appear to be less important in comparison.}
   \label{fig:overview}
\end{figure*}
\subsection{Results} 
Figure \ref{fig:overview} summarizes these analyses (one per row), with each column representing a different feature. Each panel plots the mean quoted price in euros for all profiles with a given feature value, represented on the $y$ axis, versus feature values along the $x$ axis. For example, the top left panel reports results of the \texttt{top1} analysis, depicting the average of each ``female'' profile and the average of each ``male'' profile.

Notice that results are robust across all analyses (rows) in Figure \ref{fig:overview}.\footnote{An analysis focused on median values, omitted to save space, yields equivalent results.} Age, city, car, and class have a strong effect, confirming our DOE considerations for including them. Mileage (km\_driven), on the other hand, has a weak effect, despite being reported as a powerful predictor of the number of claims at-fault \cite{lemaire2015:ua, ferreira2012:mm}.
We hypothesize that this is due to the low verifiability of this feature, which is self-reported and difficult to verify for insurance companies in the absence of a tracking device fitted on the vehicle.

Among protected features, birthplace seems to be utilized to differentiate not only between different countries, but also Italian cities, though the effects are smaller than for the previously mentioned factors.
Gender, on the other hand, seems to play a negligible role. The absence of a clear effect for this feature should not be interpreted as a guarantee that it does not directly influence actuarial models. Rather, it means that, if gender-based differences are present, they do not on average favor men or women. Section \ref{sec:discr} provides in-depth analysis on the role of gender and birthplace.

\section{Discrimination Analysis}
\label{sec:discr}

\subsection{Methods}
\label{sec:discr_meth}

In this section, we focus on the direct influence of protected attributes, i.e.\ birthplace and gender, on price. While the previous section considered average prices across feature levels, here we examine the distribution of price differences $\delta$ for pairs of profiles that differ only in one protected attribute (e.g. F-M for gender, Ghana-Milan for birthplace), which we refer to as \emph{protected pairs}. For FTU to rigorously hold, the result pages presented to protected pairs of profiles should be identical. To compensate for the effect of external factors (such as A/B testing), modest differences are deemed acceptable so long as they remain comparable to differences between two identical queries (\emph{control pairs}). Recall that protected pairs are always gathered within the same block (Figure \ref{fig:schematic}), so that the effect of time or browser session on any given protected pair should be minimal, and smaller than the effect it has on control pairs, which are gathered by definition at the very beginning and end of a block.

We conduct \texttt{top1} and \texttt{top5} analyses, collapsing each set of query results to a single price as described in Section \ref{sec:overview_meth}. Again, these analyses adopt the perspective of a driver who is only interested in the cheapest possible quote (\texttt{top1}) or a driver who is shopping around and comparing policies (\texttt{top5}). Within each analysis, we consider the vector containing price differences $\delta$ for all protected pairs with two given factor levels (e.g.\ female and male for gender).  We compute its median value $m(\delta)$ and report the $p$-value from a sign test, which tests the null hypothesis that the median difference for each pair of profiles is 0, meaning e.g.\ that we are as likely to observe a difference in favor of men as a difference in favor of women. If we reject the null hypothesis, then we are compelled to conclude that FTU does not hold, though of course failure to reject the null hypothesis does not guarantee that FTU does hold. In particular, while the condition $m(\delta)=0$ ensures that no protected group is \emph{systematically} at a disadvantage, it provides no guarantee about price differences directly determined by a protected attribute in a pair, and compensated by  a difference of opposite sign in another protected pair. To this end, we also compute the 5th and 95th percentiles (labelled $\eta_{.05}(\delta)$ and $\eta_{.95}(\delta)$, respectively), along with the percentage of protected pairs for which quote difference $\delta$ is within a tolerance threshold of 5€ ($\text{Ties}_5$). We compare these values against the ones computed for control pairs. To satisfy FTU, we would expect protected pairs and control pairs to exhibit non-zero differences with comparable frequency (summarized by $\text{Ties}_5$) and magnitude (summarized by  $\eta_{.05}(\delta)$ and $\eta_{.95}(\delta)$).

\subsection{Results}

Our numerical analysis is presented in Table \ref{tab:discr}. Rows 1-5 relate to birthplace, where Milan acts as a baseline, and positive values represent a surcharge incurred by drivers born in Rome, Naples, Romania, Ghana and Laos, respectively. Row 6 shows analogous results where the protected attribute is gender and positive differences are unfavourable for female drivers. A final row is added to summarize the effect of noise by reporting results for control pairs.

\begin{table*}[tb]
\small
  \caption{Summary of discrimination analysis. For protected pairs, we consider the vector of differences ($\delta$) in \texttt{top1} and \texttt{top5} values. We report the percentage of ties (within a 5€ tolerance threshold - $\text{Ties}_5$), the 5th and 95th percentiles ($\eta_{.05}(\delta)$,  $\eta_{.95}(\delta)$), the median difference $m(\delta)$, and the $p$-value from a sign test whose null is described in Section \ref{sec:discr_meth}. Both birthplace and gender can have a sizeable direct influence on the quotes that drivers see; the influence of birthplace is more frequent, substantial, and systematic, with drivers who are not born in Milan suffering financial disadvantages relative to Milan.}
  \label{tab:discr}
  \begin{tabular}{ccrrrrclrrrrc}
    \toprule
     & &  \multicolumn{5}{c}{\texttt{top1}} & & \multicolumn{5}{c}{\texttt{top5}}\\
    Attribute & Pairs & $\text{Ties}_5$ & $\eta_{.05}(\delta)$ &  $\eta_{.95}(\delta)$ & $m(\delta)$ & $p$ & & $\text{Ties}_5$ & $\eta_{.05}(\delta)$ &  $\eta_{.95}(\delta)$ & $m(\delta)$ & $p$ \\
    \midrule
    birthplace & Rome vs Milan & 23\% & -238 € & 207 € & 10 € & 7.6$\mathrm{e}{-08}$ & & 5\% & -202 € & 240 € & 7 € & 3.0$\mathrm{e}{-04}$\\
    birthplace & Naples vs Milan & 27\% & -60 € & 274 € &  27 € & 3.2$\mathrm{e}{-16}$ & & 6\% & -50 € & 331 € & 53 € & 7.9$\mathrm{e}{-31}$ \\
    birthplace & Romania vs Milan & 37\% & -81 € & 253 € &  17 € & 3.2$\mathrm{e}{-16}$ & & 9\% & -86 € & 225 € & 39 € & 6.8$\mathrm{e}{-27}$ \\
    birthplace & Ghana vs Milan & 30\% & -90 € & 553 € & 57 € &  7.9$\mathrm{e}{-31}$ & & 5\% & -48 € & 521 € & 84 € & 2.6$\mathrm{e}{-61}$ \\
    birthplace & Laos vs Milan & 30\% & -46 € & 312 € & 56 € &  7.9$\mathrm{e}{-31}$ & & 6\% & -60 € & 437 € & 78 € & 2.0$\mathrm{e}{-58}$ \\
    gender & F vs M & 78\% & -48 € & 127 € & 0 € & 5.3$\mathrm{e}{-02}$ & & 39\% & -173 € & 187 € & 0 € &  2.1$\mathrm{e}{-01}$ \\
    \multicolumn{2}{c}{noise control} & 93\% & -33 € & 0 € & 0 € &2.3$\mathrm{e}{-01}$ & & 89\% & -6 € & 11 € & 0 € & 5.0$\mathrm{e}{-01}$ \\
  \bottomrule
\end{tabular}
\end{table*}

Focusing on median difference $m(\delta)$, we find no systematic gender bias: the median is zero for both \texttt{top1} and \texttt{top5} analyses, with insignificant $p$-values, even before correcting for multiple hypotheses testing. However we find some sizeable price differences for gender-protected pairs, which are centered around zero, thus placing no gender at a systematic disadvantage. This finding will be discussed in the next paragraph. Birthplace, on the other hand, is used predominantly in one direction, to the advantage of drivers born in Milan. Their \texttt{top1} and \texttt{top5} average quote are consistently lower than that of foreign-born drivers from  Laos, Ghana and Romania, with median \texttt{top5} differences of 78€, 84€ and 39€, respectively. Changing birthplace from Milan to Naples also results in significantly higher quotes ($m(\delta)$ equal to 27€ for \texttt{top1} and 53€ for \texttt{top5}). Although less sizeable, drivers born in Rome also find a significant median difference with respect to their Milan-born counterparts ($m(\delta)$ equal to 10€ for \texttt{top1} and 7€ for \texttt{top5}). This  is  the  first  result  we  are aware  of  demonstrating  that  pricing  algorithms  return different quotes for drivers born in different Italian cities, even when  all  remaining factors  are  held  equal.
All $p$-values associated with birthplace are significant.

Considering the magnitude of differences directly induced by protected attributes, we find that the gender- and birthplace-based differences $\eta_{.95}(\delta)$-$\eta_{.05}(\delta)$ in the \texttt{top5} results range from 311€ to 569€, compared with a value of 17€ for control pairs. The frequency of $\text{Ties}_5$ for \texttt{top5} is below 10\% for all birthplace-protected pairs and below 40\% for gender-protected pairs, compared against a value of 89\%  for control pairs. Similar if somewhat weaker patterns obtain in the \texttt{top1} results. We interpret these findings as evidence that gender and birthplace have a direct and substantial influence in the result pages of this comparison website. Histograms for these differences are reported in Appendix A of the auxiliary material.

In sum, the pricing algorithms generating the RCA quotes that drivers obtain through this popular aggregator violate FTU: when all else is held constant, both gender and birthplace have sizable effects on the quoted prices, even though, in the case of gender, the direction of this effect is not systematic, i.e. the median effect is 0. Given that aggregators have become a primary point of access to RCA subscription, these results point to potentially nontrivial violations of existing laws and regulations.

It is not immediately clear to what extent these results arise from individual companies’ pricing algorithms vs. the behavior of the aggregator. In this regard, we note that 4 out of 9 of the companies in the results do not appear to use gender or birthplace directly for pricing insurance. This suggests (1) that these results are probably not due to the aggregator alone and (2) that the use of gender and birthplace does not qualify as a fundamental business need, which might otherwise partially explain violations of FTU.  While the aggregator may in theory offer different prices than are offered on companies' own websites, studies of prevalent business models for aggregators suggest the contrary \cite{ivass2014:aggregator,mckinsey2018}. To investigate whether the pricing patterns we find are independent of the aggregator, in Section 7 we analyze a dataset gathered from a single insurance company’s website, comparing these quotes against the ones obtained on the aggregator.

\section{Output Variability}
\label{sec:target}

\subsection{Methods}

In this section we analyze the effect of each factor included in our DOE on the frequency $f_q$ with which insurance companies appear in  quotes for specific profiles. We report the results from four companies listed in Table \ref{tab:summ_data}, for which $f_q$ displays a clear dependency on one or more factors. We also aggregate these results from the perspective of users, detailing how different features affect the average number of quotes they see. 

\subsection{Results}

\begin{figure*}[!tb]
    \centering
    \includegraphics[width=\linewidth]{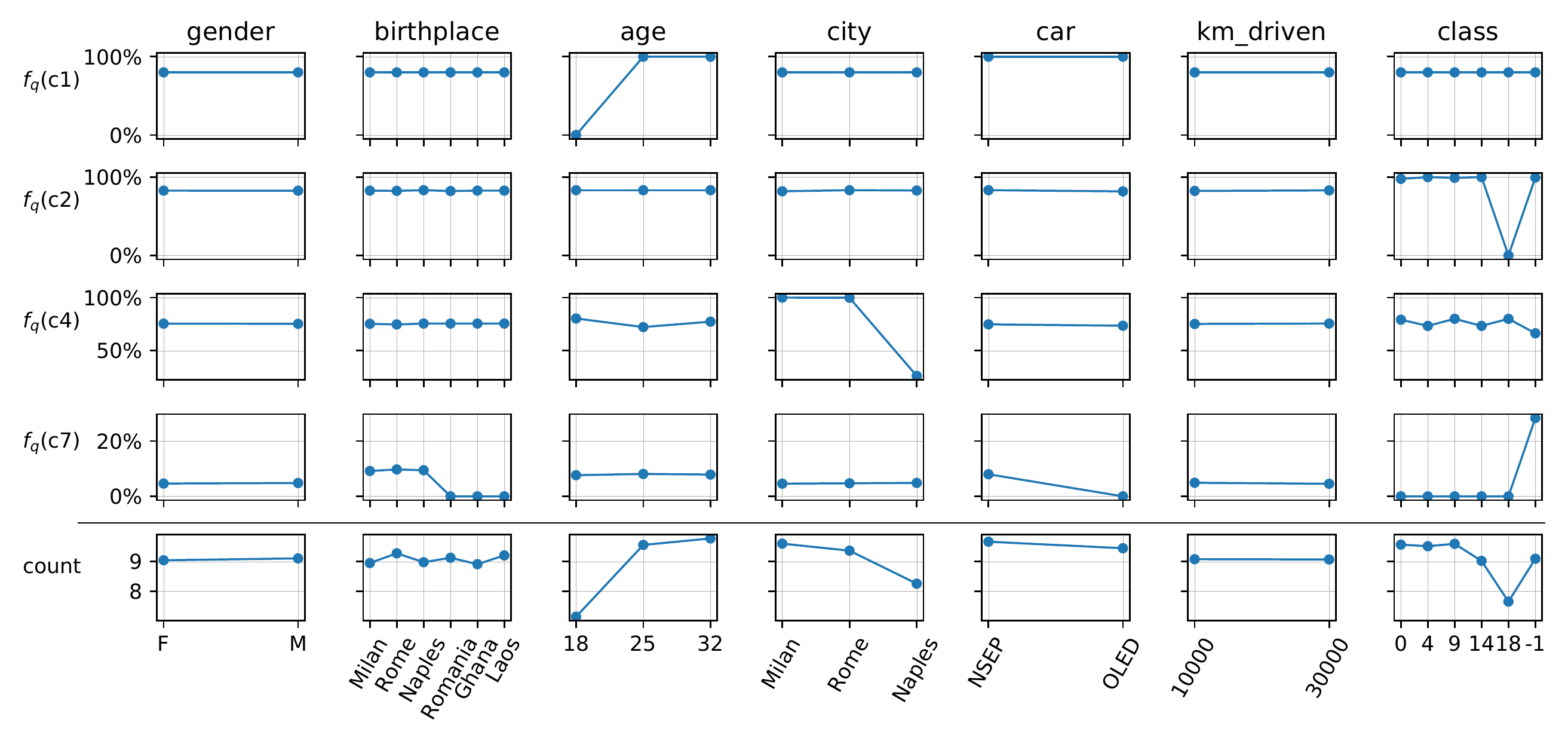}
  \caption{Output variability. Influence of each factor on frequency of appearance in result pages for company \texttt{c1} (row 1), \texttt{c2} (row 2), \texttt{c4} (row 3), \texttt{c7} (row 4) and total number of quotes (row 5). Strong patterns are visible for factors age, city and class.}
   \label{fig:num_quotes}
\end{figure*}

Figure \ref{fig:num_quotes} contains a summary of our results, where each column represents a factor, with all its possible values on the $x$ axis.
Rows 1-4 depict $f_q$ for c1, c2, c4 and c7 respectively. Interesting patterns emerge over the 2160 profiles that were tested. Company c1 is never present in result pages for 18-year-old drivers. Company c2 is absent from result pages for drivers in the worst BMS class (18). Both these results are very strong, as c1 and c2 are otherwise present 100\% of the time. Company c4 is always present in result pages for residents of Rome and Milan, but its frequency of appearance drops to 26\% for Naples. Company c7, appearing in only 4\% of result pages, seems to focus on Italian-born drivers of non-OLED cars with no claim history.

Row 5 of Figure \ref{fig:num_quotes} aggregates these results from the perspective of users, detailing how different features affect the average number of quotes they see, reported on the $y$ axis. 
Age plays a major role, with 18- and 32-year-olds seeing on average 7.1 and 9.8 quotes respectively. Municipality is also important: more quotes are available for residents of Milan than for residents of Naples. Claim history is another important factor, with drivers in BMS classes above 9 seeing fewer quotes than drivers with more favorable classes. Overall, these are also factors which have a strong influence on insurance premiums, as depicted in Figure \ref{fig:overview}. Profiles perceived as risky see fewer, more expensive quotes. 

Just like price, the number of quotes may be subject to noise, due e.g.\ to A/B testing or technical issues. We quantify this effect by considering control pairs. We notice that 17\% of result page pairs differ by 1 in the number of quotes returned, resulting in an average absolute difference of 0.17 quotes for identical profiles. We regard this figure as an estimate of noise affecting the number of quotes returned by the aggregator in its result pages. As shown in the bottom row of Figure \ref{fig:num_quotes}, age, city and class induce systematic differences, one order of magnitude larger than this threshold.

To illustrate the potential impact of these findings on drivers, let us consider matching profiles that differ only for age, and let us pair 18-year-olds with their 32-year-old counterparts. In 26\% of the resulting pairs, the company providing the cheapest quote to the 32-year-old driver is absent from the result page of the matching 18-year-old. This clearly reduces the opportunities available to some younger drivers, hiding from them potentially favourable premiums, which in turn can contribute to an increase in their expenses. This problem is relevant also for factors that are not associated with systematic output variability. Focusing on gender, for example, if we consider gender-protected pairs such that $\delta_{\texttt{top1}} > \eta_{.95}(\delta_{\texttt{top1}})$, i.e. the pairs with most extreme \texttt{top1} differences in favour of men (top 5 percentiles), we find that the company providing the cheapest quote to the male profile is absent from the result page of his female counterpart 84\% of the time. In other words, the most extreme differences in \texttt{top1} price for gender-protected pairs appear to be caused by output variability in the aggregator result pages.

\section{Aggregator Influence on Premiums}
\label{sec:aggregator_influence}

\subsection{Methods}
Based on reports on aggregators and their typical business model, we expect their influence on quoted prices to be null or negligible \cite{ivass2014:aggregator,mckinsey2018}. In this Section we verify that the key pricing trends obtained on the comparison website are also present on an individual company website. Considering a single company and a single product (\texttt{c1/a}), we repeat our data collection procedure, with doubly-nested randomization and control (summarized in Figure \ref{fig:schematic}), directly on the company website. We concentrate on a subset of our dataset, comprising 32-year-old drivers with BMS classes 0 and 4. We choose this subset since (1) \texttt{c1/a} is always present in the respective aggregator result pages, allowing for a direct comparison; and (2) this is the most representative subset in our sample, as very young drivers and BMS classes above 4 are quite rare among Italian RCA subscribers \cite{ivass2020q1,ivass2020q2}. The resulting dataset consists of 288 regular quotes and 24 control quotes, gathered in the second half of December, 2020.

Concentrating on \texttt{c1/a}, we mimic the analyses from Sections \ref{sec:overview} and \ref{sec:discr}, i.e.\ an overview of the most important factors and a discrimination analysis focused on protected pairs. As the dataset collected from the aggregator predates this one by six months, we do not attempt a rigorous characterization of the comparison website effect in terms of fees and discounts. Instead, we are mainly interested in evaluating whether the key trends from Sections \ref{sec:overview} and \ref{sec:discr} are confirmed.

\subsection{Results}
\begin{figure*}[!thb]
    \centering
    \includegraphics[width=\linewidth]{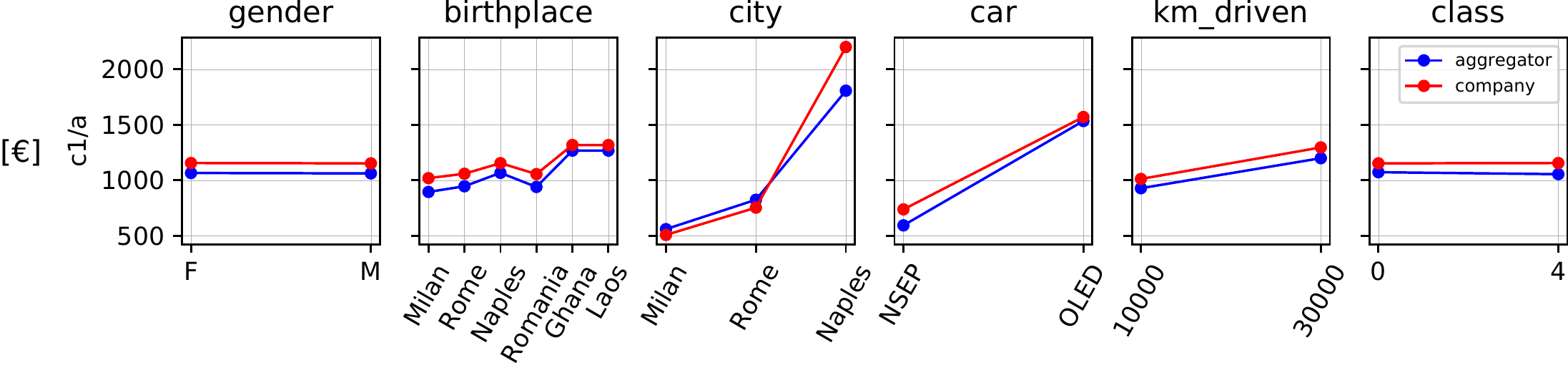}
  \caption{Consistency of general trends in \texttt{c1/a} pricing across the two datasets. Each panel represents a different factor, with its levels on the $x$ axis. The $y$ axis depicts the mean price for all profiles with a given factor level, gathered on the aggregator (blue) and company website (red). Trends are stable across both datasets, despite having been collected six months apart.}
   \label{fig:aggr_vs_comp}
\end{figure*}

Figure \ref{fig:aggr_vs_comp} depicts the effect of each factor, as an average price for profiles sharing a factor level, across each of the remaining factors, similarly to Figure \ref{fig:overview}. 
Overall trends are confirmed for every factor across all levels. 

Table \ref{tab:discr_aggr_vs_comp} reports summary statistics for protected pairs on both datasets; corresponding histograms are available in Appendix B of the auxiliary material.
\begin{table*}[tb]
\small
\centering
  \caption{Consistency in birthplace- and gender-related trends in \texttt{c1/a} pricing across the two datasets. Differences in price for protected pairs ($\delta$). Rows are consistent with Table \ref{tab:discr}. Columns report the percentage of ties within a 5€ tolerance threshold ($\text{Ties}_5$), 5th, 95th percentile and median difference ($\eta_{.05}(\delta)$, $\eta_{.05}(\delta)$ and $m(\delta)$), as computed from the dataset gathered on the aggregator (Aggr.) and directly on the company website six months later (Comp.). Trends are stable across both datasets.}
  \label{tab:discr_aggr_vs_comp}
  \begin{tabular}{cccrrrrrrrrrr}
    \toprule
     & & \multicolumn{2}{c}{$\text{Ties}_5$} & & \multicolumn{2}{c}{$\eta_{.05}(\delta)$} & & \multicolumn{2}{c}{$m(\delta)$} & & \multicolumn{2}{c}{$\eta_{.95}(\delta)$}\\
     Attribute & Pairs & Comp. & Aggr. & & Comp. & Aggr. & & Comp. & Aggr. & & Comp. & Aggr. \\
     \midrule
     birthplace & Rome vs Milan & 29\% & 17\%  &  & -1 € & -1 € & & 16 € & 23 € &  & 121 € & 197 €  \\
     birthplace & Naples vs Milan & 20\% & 17\%  &  & -1 € &  -1 € & & 144 € & 143 € &  & 294 € & 435 €  \\
     birthplace & Romania vs Milan & 42\% & 38\%  &  & -6 € & -10 € & & 0 € & 6 € &  & 124 € & 211 €  \\
     birthplace & Ghana vs Milan & 8\% & 0\%  &  & 0 € & 92 € & & 213 € & 243 € &  & 1116 € & 867 €  \\
     birthplace & Laos vs Milan & 8\% & 0\%  &  & 0 € & 87 € & & 213 € & 243 € &  & 1116 € & 914 €  \\
     gender & F vs M & 80\% & 76\%  &  & 0 € & -5 € & & 0 € & 0 € &  & 14 € & 19 € \\
     \multicolumn{2}{c}{noise control} & 96\% & 100\%  &  & 0 € & 0 € & & 0 € & 0 € &  & 0 € & 0 € \\
  \bottomrule
\end{tabular}
\end{table*}
Similarly to Table \ref{tab:discr}, we consider both birthplace- and gender-protected pairs in rows 1-6, with a final row focused on control pairs. In each row, we report the frequency of ties within a 5€ tolerance threshold ($\text{Ties}_5$), along with the median, 5th and 95th percentiles, labelled $m(\delta)$, $\eta_{.05}(\delta)$ and $\eta_{.95}(\delta)$ respectively.

Overall we find stable trends across both datasets, as summarized in Table \ref{tab:discr_aggr_vs_comp}.
\begin{itemize}
    \item About 80\% of gender-protected pairs are tied. Ties are less frequent between birthplace-protected pairs within the EU (17\%-42\%) and very rare when comparing drivers born in Milan with their counterparts born in Ghana or Laos (0\%-8\%).
    \item $\eta_{.05}(\delta)$ is weakly (if at all) negative, showing that the baseline factor level (Milan for birthplace, male for gender) is rarely at a disadvantage.
    \item  $m(\delta)$ is similar in both datasets, confirming a systematic and sizeable financial disadvantage for drivers born in Naples, Ghana and Laos ($m(\delta) >$ 100€).
    \item $\eta_{.95}(\delta)$ is always larger than 100€ for birthplace-protected pairs, reaching a 1,000€ surcharge for Ghana and Laos.
    \item noise control shows minimal differences for identical queries.
\end{itemize}

In sum, these results show that the effect of the comparison website on the prices quoted in its result pages (if any) is modest in comparison with the effect of pricing algorithms employed by company \texttt{c1}. As a final remark, it is worth highlighting the strong financial disadvantage measured for Laos-born drivers despite the small number of Laos citizens residing in Italy \cite{istat2020} and available to company \texttt{c1} to infer the ``effect'' of this feature in risk models.\footnote{In this case, we can likely rule out that the feature is being used as a proxy for the country where drivers learned to drive, since the company website explicitly queries the year of arrival in Italy, and our input, 2004, predates by 2 years the driver's license issue date.}

\section{Discussion and Conclusion}
\label{sec:concl}

We have conducted an audit of algorithms in the Italian car insurance market, gathering quotes from a widely-used comparison website, to answer the following questions about pricing and access.

\textbf{RQ1: What are the factors that play a major role in setting RCA premiums?}\\
We examined the prices stratified on each feature, averaged across each of the remaining factors. We found that driver age, city, vehicle and claim history are important factors for RCA pricing, at least for the sample we considered. Contrary to our expectation, the levels we tested for mileage led to small average price differences, probably due to the low verifiability of this feature. Birthplace and gender also induced smaller average fluctuations, which we analyzed more in detail in light of their sensitive nature and existing legislation against their direct use.

\textbf{RQ2: Do gender and birthplace directly influence quoted premiums?}\\
Both factors have a direct influence on the quotes offered to users: we paired driver profiles, so that they only differ for gender or birthplace, and found that quotes provided to them vary frequently and substantially. These differences are larger than those present in control (identical) pairs.

More in detail, we analyzed the distribution of paired differences, finding that gender-related differences are centered around zero, confirming the finding for \textbf{RQ1} that no gender is systematically at a disadvantage. However, some sizeable differences were measured in both directions, showing that gender can have a direct non-negligible influence on quoted price.  Birthplace-related differences, on the other hand, exhibit patterns of systematic discrimination. Foreign-born drivers and natives of Naples are consistently charged more expensive premiums when compared against drivers born in Milan, \emph{ceteribus paribus}.
We interpret these findings as a violation of Fairness Through Unawareness (FTU), which is the fairness principle that (most closely) aligns with European legislation on gender equality in insurance \cite{ecj2011, eu2012:guidelines} and Italian soft regulation against nationality-based discrimination \cite{unar2012,ivass2014}. We repeated our data collection procedure on a single company's website, focusing on the most representative subset of our sample. Comparative analysis supports the key trends discussed above, confirming that the influence of the aggregator on quoted prices is moderate, if any.

\textbf{RQ3: Do riskier driver profiles see fewer quotes on comparison websites?} \\
We analyzed the frequency with which insurers appear in result pages for different profiles, finding that some companies are systematically absent from result pages for certain driver segments. In sum, 18-year-olds, drivers with a bad claim history, and residents of Naples appear to be the least desirable categories in our dataset: when they query the comparison website, they end up receiving, on average, fewer quotes.
These results are compatible with anecdotal findings from IVASS on aggregator output variability, associating riskier profiles with fewer RCA quotes \cite{ivass2014:aggregator}. The evidence we found on our medium-size sample represents a confirmation that strategic choices seem to be in place, providing users of comparison websites with unequal opportunity and access to products based on their risk profile.

\noindent \textbf{Limitations and future work.} Our analyses hinge on quotes for 2,160 driver profiles, a dataset of limited size and not fully representative of the Italian driving population at large. Moreover, we were only able to examine a subset of the relevant features, which does not fully characterize the behavior of the pricing algorithm. While our experiments show a violation of FTU, we did not attempt to quantify the impact of the discrimination that we uncovered on Italian society at large. This would be a large and complex endeavor, and an interesting target for future work.

\begin{acks}
We are indebted to Tomislav Jajčević and Chiara Maltese for early discussion about this work.
Part of this work was supported by MIUR (Italian Minister for Education) under the initiative "Departments of Excellence" (Law 232/2016).
\end{acks}


\DeclareRobustCommand\ECJlongname{ - European Court of Justice}
\DeclareRobustCommand\EUlongname{ - European Union}
\DeclareRobustCommand\EClongname{ - European Commission}
\DeclareRobustCommand\DGMOVElongname{ - EU Directorate-General for Mobility and Transport}
\DeclareRobustCommand\IVASSlongname{ - Institute for the Supervision of Insurance}
\DeclareRobustCommand\UNARlongname{ - National Anti-Racial Discrimination Office}
\DeclareRobustCommand\ASGIlongname{ - Association for Judicial Studies on Immigration}
\DeclareRobustCommand\UNRAElongname{ - Unione Nazionale Rappresentanti Autoveicoli Esteri}

\bibliographystyle{ACM-Reference-Format}
\balance
\bibliography{biblio}

\clearpage

\begin{screenonly}
\appendix
\section{Price differences for protected pairs in comparison website}
\label{app:hist}
We report histograms for price differences quoted to protected pairs of profiles in Figure \ref{fig:hist}. Rows 1-5 depict, along the $x$ axis, the surcharge incurred by drivers born in Rome, Naples, Romania, Ghana and Laos (respectively) when compared against their counterparts born in Milan. Row 6 depicts differences in price for gender-protected pairs. Positive values represent a financial advantage for the baseline (Milan for birthplace, male for gender).

Birthplace clearly plays an important role, with differences strongly skewed towards positive values, signaling a systematic bias in favour of drivers born in Milan. Gender is used less frequently and in a more balanced fashion, but can still determine sizeable differences for aggregator users. 

\section{Consistency of trends based on protected attributes on company website and aggregator}
\label{app:hist2}
Figure \ref{fig:hist2} depicts histograms for price differences quoted to protected pairs of profiles for the insurance product labelled \texttt{c1/a}, reporting both the price obtained on the aggregator (blue) and the price obtained on the company website (orange). Rows are consistent with Figure \ref{fig:hist}. These results pertain to a subset of the full sample, as described in Section \ref{sec:aggregator_influence}. Despite the fact that the aggregator dataset predates the company dataset by six months,  key trends are stable.

\begin{figure}[H]
    \centering
    \includegraphics[width=\linewidth]{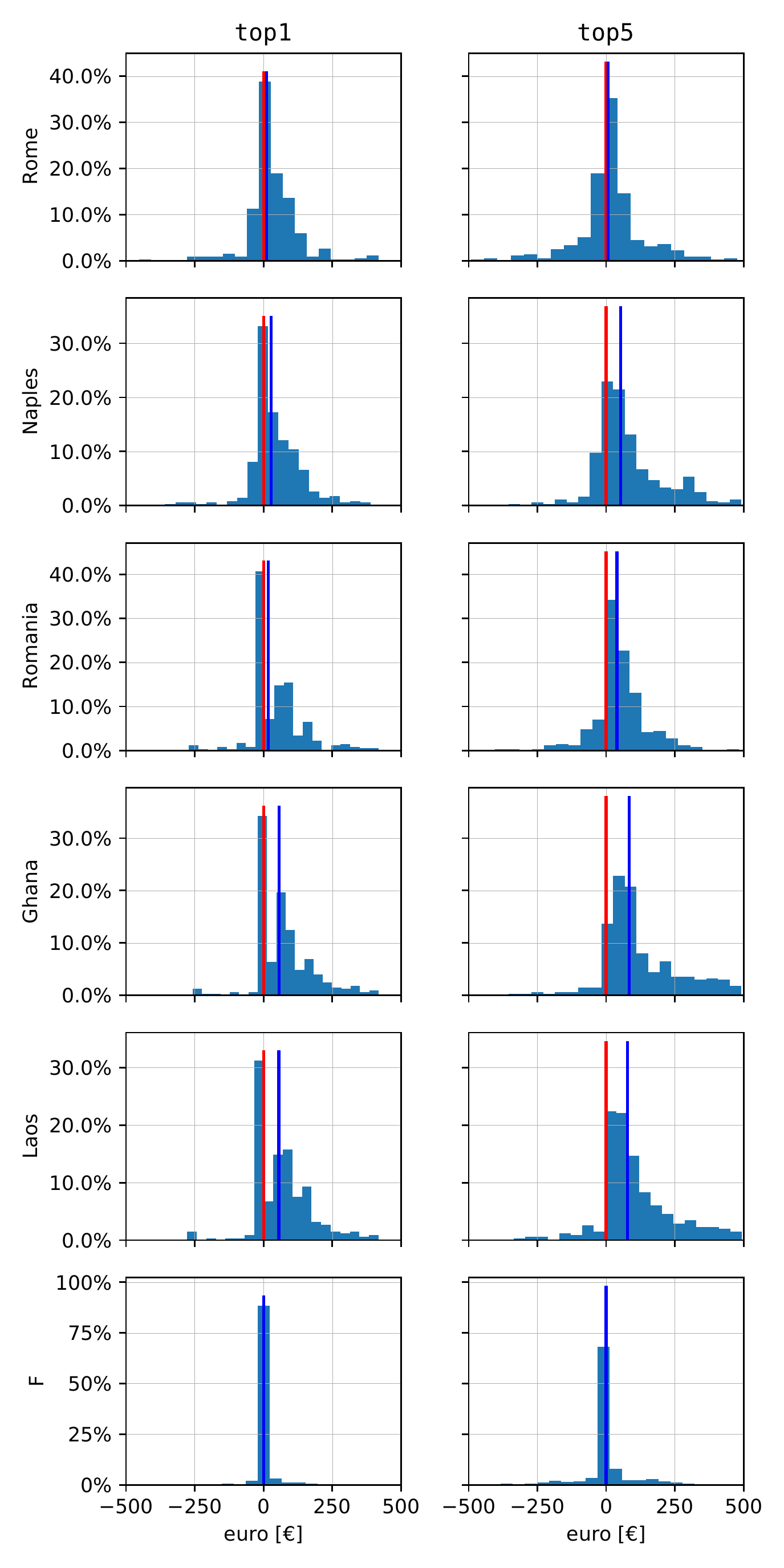}
  \caption{Birthplace- and gender-based discrimination. Histogram of paired differences in cheapest quote (col. 1) and average of 5 cheapest quotes (col. 2). Rows 1-5 focus on birthplace, with Milan as a baseline, depicting paired differences with respect to Rome (row 1), Naples (row 2), Romania (row 3), Ghana (row 4), Laos (row 5). Positive means Milan is cheaper.  Row 6 depicts gender-based differences (F-M). Vertical blue lines represent the median difference, while red lines are the median difference between regular and control quotes (zero).  The $x$ axis is clipped  between \mbox{-500} and 500 €. Both birthplace and gender can have a sizeble influence, the former being more frequent, strong and systematic in one direction.}
   \label{fig:hist}
\end{figure}

\begin{figure}[H]
    \centering
    \includegraphics[width=\linewidth]{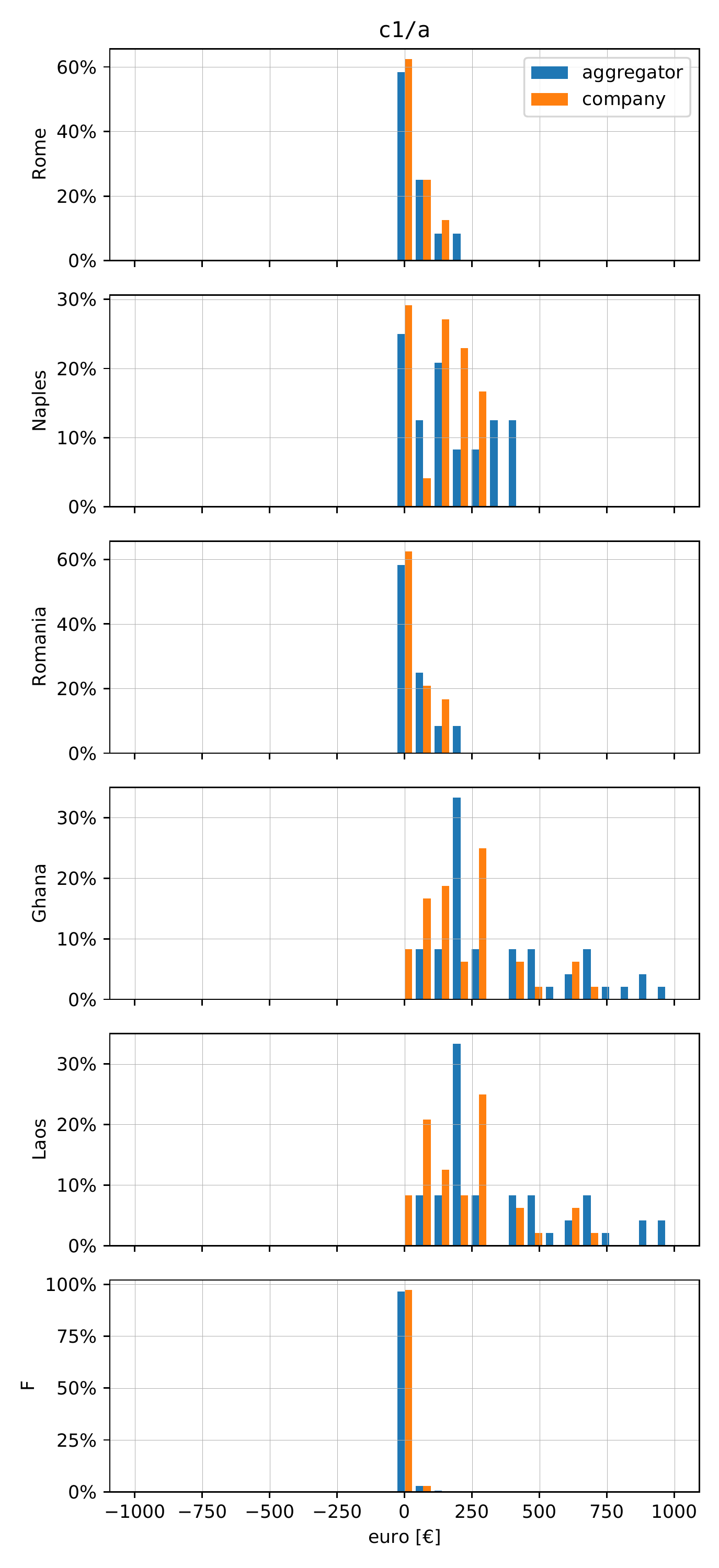}
  \caption{Consistency of trends for birthplace- and gender-based discrimination on company website and aggregator. Histogram of differences in \texttt{c1/a} quote provided on aggregator (blue) and on company website (orange) to different protected pairs. Rows are consistent with Figure \ref{fig:hist}. The $x$ axis is clipped  between -1000 and 1000 €. Key trends are stable, despite the fact that datasets have been collected six months apart.}
   \label{fig:hist2}
\end{figure}

\end{screenonly}

\end{document}